# Unlocking Health Insights with SDoH Data: A Comprehensive Open-Access Database and SDoH-EHR Linkage Tool

Zhenhong Hu[a,b,‖], Esra Adiyeke[a,b,‖], Ziyuan Guan[a,b], Divya Vellanki[a,b], Jiahang Yu[a,b], Ruilin Zhu[a,b], Yuanfang Ren[a,b], Yingbo Ma[a,b], Annanya Sai Vedala[a,b], Tezcan Ozrazgat-Baslanti[a,b,*], Azra Bihorac[a,b,*]

‖ These first authors have contributed equally.

* These senior authors have contributed equally.

[a] Intelligent Clinical Care Center, University of Florida, Gainesville, FL, USA

[b] Department of Medicine, Division of Nephrology, Hypertension, and Renal Transplantation, University of Florida, Gainesville, FL, USA

**Corresponding author:** Azra Bihorac MD, MS, Division of Nephrology, Hypertension, and Renal Transplantation, Department of Medicine, PO Box 100224, Gainesville, FL 32610-0224. Telephone: (352) 294-8580; Fax: (352) 392-5465; Email: abihorac@ufl.edu

## Abstract

Background: Social determinants of health (SDoH) play a crucial role in influencing health outcomes, accounting for nearly 50% of modifiable health factors and bringing to light critical disparities among disadvantaged groups. Despite the significant impact of SDoH, existing data resources often fall short in terms of comprehensiveness, integration, and usability. To address these gaps, we developed an extensive Exposome database and a corresponding web application, aimed at enhancing data usability and integration with electronic health record (EHR) to foster personalized and informed healthcare.

Methods: We created a robust database consisting of a wide array of SDoH indicators and an automated linkage tool designed to facilitate effortless integration with EHR. The database architecture focuses on scalability, robust spatial relationships, and optimized query performance. We also emphasized the development of a user-friendly interface to cater to researchers, clinicians, and public health professionals. Future upgrades of the database will include additional SDoH indicators and the incorporation of longitudinal data for trend analysis.

Results: The resultant Exposome database and web application offer an extensive data catalog with enhanced usability features. The automated linkage tool has demonstrated efficiency in integrating SDoH data with EHRs, significantly improving data accessibility. Initial deployment has confirmed scalability and robust spatial data relationships, facilitating precise and contextually relevant healthcare insights.

Conclusion: The development of an advanced Exposome database and linkage tool marks a significant step toward enhancing the accessibility and usability of SDoH data. By centralizing and integrating comprehensive SDoH indicators with EHRs, this tool empowers a diverse range of users to access high-quality, standardized data. Continued efforts will focus on incorporating additional indicators, ensuring data quality, and promoting global collaborations to support health equity. This resource promises lasting impact on personalized healthcare, fostering an informed and equitable health landscape.

## Introduction

Social determinants of health (SDoH) refer to the wide range of conditions influencing individuals' health that extend beyond lifestyle choices or healthcare access.[1] Together with SDoH impacted by the environment in which a person lives, individual-level health-related social needs are key factors that affect health distribution and outcomes.[2,3] Research indicates that the majority of modifiable drivers of health were attributable to SDoH, illustrating a social gradient in health where those with fewer socioeconomic resources, i.e. disadvantaged SDoH, face greater barriers in achieving optimal health.[4,5] Specifically, area-level SDoH indicating worsened socioeconomic levels were associated with higher mortality risks[6-8] and major adverse cardiovascular events[9]. These adverse conditions correlated with longer hospital stay[10] increased intensive care unit admission[11], and greater likelihood for postoperative complications.[12] Considering the manifested impact of SDoH on health and the substantial difference in SDoH by neighborhood, identifying individuals' SDoH can guide research and aid in developing tailored approaches that address specific needs and targeted intervention. Recognizing this, the United States (US) Department of Health and Human Services designated SDoH as a central focus area in the Healthy People 2030 framework outlining the agenda and national goals for improving health and well-being.[13]

Information required for deriving area-level SDoH is gathered through surveys, data aggregation from public sources and routine monitoring activities performed by local or national bodies and organizations.[14] This creates a rich landscape of SDoH data resources that are spread across different platforms and generated for various geographic units like county or census tracts. To address challenges in identifying and processing the SDoH data, several SDoH data sets from different public resources were coalesced into databases or platforms to facilitate access and use.[4,14-18] While being used for research and policy generation purposes, these resources primarily depend on

American Community Survey of Census Bureau and do not comprehensively cover the majority of the public sources.

Inferring an individual's SDoH typically involves making estimations based on their place of residence. Geocoding, which converts address data to geographic coordinates and links these coordinates to administrative boundaries like ZIP codes, enables the mapping of individuals' locations to area-based SDoH.[19] Currently, there are offline and online tools for geocoding large sets of address data. Offline geocoding tools like ESRI ArcGIS, SAS Geocoder and DeGAUSS, have been in use for several years; however, they are often commercial products with high costs or require complex technical skills to use.[20] Online tools like Google Maps involve sharing address information with the service provider, which can raise privacy concerns since address and census-level data are classified as sensitive information.[21]

The variety in characteristics, focus areas, and providers of SDoH resources, coupled with the technical and privacy challenges associated with geocoding address data, underscores the necessity of eliminating barriers for policymakers, healthcare professionals, and researchers to effectively navigate and utilize these data sources. To address the critical need for a unified platform that streamlines a comprehensive collection of publicly available SDoH resources with a geocoding scheme capable of offline processing of large batches of address data, while preserving information privacy, we developed a unified platform and web application called Exposome (https://exposome.rc.ufl.edu/). This platform integrates diverse SDoH and environmental databases from various providers. e have also shared an offline open-source geocoding toolkit (https://github.com/bihorac-LAB/Exposome) with the research community.

## Building the Exposome database

The recognition of the crucial role that SDoH and environmental factors play in influencing clinical outcomes has intensified the demand for integrated datasets among researchers globally. Despite the availability of diverse data sources from government agencies and research institutions, the field lacks a centralized, standardized database that facilitates efficient data access, visualization, and extraction.

Addressing this gap, we developed a comprehensive spatial database using PostgreSQL with the PostGIS extension, drawing on design principles from the Observational Health Data Sciences and Informatics (OHDSI) GIS working group. Our database architecture emphasizes fundamental spatial relationships and standardized variable linkages. Given the inherently geographic nature of SDoH and environmental data, we integrated hierarchical geographic boundaries—including Census Tracts, ZIP Code Tabulation Areas (ZCTAs), Counties, and States—as foundational spatial layers to establish robust relationships between variable tables and spatial geometries.

Census Tract identifiers (FIPS codes) serve as the primary geographic unit for most SDoH datasets. To facilitate seamless data integration, we developed specialized geocoding toolkits that enable investigators to convert source addresses into both coordinate pairs and FIPS codes. To preserve patient privacy and comply with data protection regulations, these geocoding operations are performed locally using the DeGauss framework. Our implementation supports both standalone file processing and direct data extraction from OMOP-formatted databases, ensuring compatibility with existing clinical data infrastructures.

## Data Sources

Our database's foundation rests on meticulously curated data sources from authoritative organizations. We prioritized datasets based on their reliability, update frequency, documentation quality, and widespread use in research communities. The selection process involved a systematic evaluation of each dataset's methodological rigor, completeness, and integration potential with existing clinical research frameworks.

**Table 1. SDOH and environmental dataset characters**

| Dataset Name | Data Source Organization | Variable Count | Spatial Scale | Domains Covered |
|---|---|---|---|---|
| Social Determinants of Health (SDOH)Database | Agency for Healthcare research and Quality (AHRQ) | 405 | Census tract | SDoH |
| Social Vulnerability Index (SVI) | Centers for Disease Control (CDC) | 158 | Census tract | SDoH |
| Environmental Justice Index (EJI) | Centers for Disease Control (CDC) | 117 | Census tract | SDoH |
| Area Deprivation Index (ADI) | Neighborhood Atlas Area Deprivation Index (ADI) | 4 | Census tract | SDoH |
| DistrictsFullList New | HHS Protect Public Data Hub | 70 | Census tract | SDoH |
| COVID-19 Reported Patient Impact and Hospital Capacity by Facility | U.S. Department of Health & Human Services | 128 | County | SDoH |
| COVID-19 Diagnostic Lab Testing | HHS Protect Public Data Hub | 9 | State | SDoH |
| Hospital_Data_Coverage_Report_og | HHS Protect Public Data Hub | 147 | Census Tract | SDoH |
| Area Health Resources Files Diversity Dashboard Data | Health Resources & Services Administration (HRSA) | 38 | state | SDoH |
| Area Health Resources Files (county) | Health Resources & Services Administration (HRSA) | 4306 | county | SDoH |
| Area Health Resources Files (State and National) | Health Resources & Services Administration (HRSA) | 1432 | state | SDoH |
| Crime in the United States, by State | Uniform Crime Reporting (UCR) | 84 | state | SDoH |
| Food Environment Atlas | United States department of agriculture (USDA) | 293 | County | SDoH |
| Food Access Research Atlas | USDA Food Access Research Atlas (FARA) | 147 | Census Tract | SDoH |
| National Walkability Index | United States Environmental Protection Agency (EPA) | 117 | Census block group | SDoH |
| Social Deprivation Index (SDI) | Robert Graham Center (RGC) | 18 | Census Tract | SDoH |
| Child Opportunity Index (COI) Child population data | diversitydatakids.org | 8 | Census Tract | SDoH |
| Child Opportunity Index (COI) COI 3.0 overall index and three domains | diversitydatakids.org | 38 | Census Tract | SDoH |
| Child Opportunity Index (COI) subdomains | diversitydatakids.org | 128 | Census Tract | SDoH |
| Rural-Urban Commuting Area Codes | U.S. Department of Agriculture | 6 | Census Tract | SDoH |
| Poverty Area Measures | U.S. Department of Agriculture | 24 | Census Tract | SDoH |
| Ozone (daily) | United States Environmental Protection Agency (EPA) | 28 | Latitude/Longitude | Environment |
| CO (daily) | United States Environmental Protection Agency (EPA) | 28 | Latitude/Longitude | Environment |
| NO2 (daily) | United States Environmental Protection Agency (EPA) | 28 | Latitude/Longitude | Environment |
| SO2 (daily) | United States Environmental Protection Agency (EPA) | 28 | Latitude/Longitude | Environment |
| PM2.5 FRM/FEM Mass (88101) (daily) | United States Environmental Protection Agency (EPA) | 28 | Latitude/Longitude | Environment |
| PM2.5 non FRMFEM Mass (88502) (daily) | United States Environmental Protection Agency (EPA) | 28 | Latitude/Longitude | Environment |
| PM2.5 Speciation(daily) | United States Environmental Protection Agency (EPA) | 28 | Latitude/Longitude | Environment |
| PM10 Mass (81102) (daily) | United States Environmental Protection Agency (EPA) | 28 | Latitude/Longitude | Environment |
| PM10 Speciation(daily) | United States Environmental Protection Agency (EPA) | 28 | Latitude/Longitude | Environment |
| PMc Mass (86101) (daily) | United States Environmental Protection Agency (EPA) | 28 | Latitude/Longitude | Environment |
| Barometric Pressure (64101) (daily) | United States Environmental Protection Agency (EPA) | 28 | Latitude/Longitude | Environment |
| RH and Dewpoint(daily) | United States Environmental Protection Agency (EPA) | 28 | Latitude/Longitude | Environment |
| Temperature (62101) (daily) | United States Environmental Protection Agency (EPA) | 28 | Latitude/Longitude | Environment |
| Winds (Resultant)(daily) | United States Environmental Protection Agency (EPA) | 28 | Latitude/Longitude | Environment |
| HAPs(daily) | United States Environmental Protection Agency (EPA) | 28 | Latitude/Longitude | Environment |
| Lead(daily) | United States Environmental Protection Agency (EPA) | 28 | Latitude/Longitude | Environment |
| NONOxNOy(daily) | United States Environmental Protection Agency (EPA) | 28 | Latitude/Longitude | Environment |
| VOCs(daily) | United States Environmental Protection Agency (EPA) | 28 | Latitude/Longitude | Environment |
| Daily AQI by County | United States Environmental Protection Agency (EPA) | 8 | County | Environment |
| Daily AQI by CBSA | United States Environmental Protection Agency (EPA) | 8 | CBSA | Environment |

These data sources were selected for their robust methodological foundations, comprehensive documentation, and established utility in health research. To compile all the datasets mentioned above, we employed a combination of manual downloads and automated methods. For certain datasets, we accessed publicly available repositories and manually downloaded the files through provided links. This process involved visiting the website, navigating to the appropriate sections, and downloading the data in formats such as CSV or Excel files. For other datasets such as TIGER Census Tract Data, we utilized APIs to streamline the data retrieval process. By integrating these APIs into our system, we were able to programmatically request and download the data, ensuring that we always had the most up-to-date information. This combination of manual and automated methods allowed us to efficiently gather comprehensive datasets while maintaining data accuracy and timeliness.

### Database Structure

Implementing a comprehensive SDoH and environmental data repository required a robust yet flexible database architecture. We developed a scalable PostgreSQL-based system with PostGIS extension, incorporating spatial capabilities while maintaining efficient data management and retrieval mechanisms.

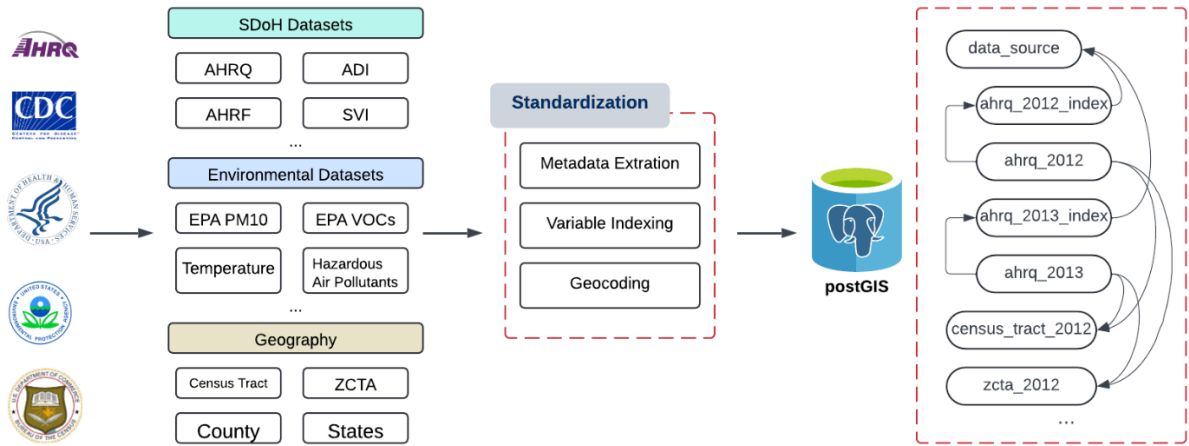

**Figure 1. Overview of the Exposome database.**

As shown in Figure 1, data sources are processed through a standard procedure to extract the key information, transform the variables and then load them into the database structures. The entire process is similar to an ETL pipeline.

The database architecture employs a modular design composed of several key components.

  a. Data Source Registry: This centralized metadata repository tracks all integrated datasets, including both SDoH/environmental and geometric data sources. Essential metadata elements, such as unique source identifiers, spatial boundary definitions, temporal coverage ranges, and access permissions, are meticulously maintained.
  b. Variable Index Tables: Automatically generated tables catalog metadata for each dataset's variables. They standardize variable information through detailed descriptions, unit specifications, and value format constraints, incorporating OHDSI CDM-compatible concept mappings and standard terminology alignments.
  c. Value Storage Tables: Dedicated tables maintain actual measurements and observations, linking them to appropriate geometric identifiers while preserving temporal references and quality indicators. This design ensures efficient data retrieval and maintains relationships between values and their spatial contexts.
  d. Geometric Reference Framework: Supports multiple resolution levels, from Census Tract FIPS codes to ZIP Code Tabulation Areas (ZCTA), county boundaries, and state demarcations. For environmental datasets requiring higher spatial precision, the system accommodates coordinate-based references, facilitating comprehensive spatial analysis capabilities.

### Implementation Considerations

Our architecture is designed for scalability and optimized query performance, making it suitable for future dataset integration and diverse research workflows. Key implementation considerations include maintaining robust spatial relationships and data integrity across linked tables, allowing for both standalone operations and integration with established clinical data repositories through standardized interfaces. This approach ensures efficient data retrieval and offers flexibility for various research applications.

Specifically, the database schema covers key variables such as income, education, housing, employment, and environmental factors. It spans multiple geographic granularity levels (ZIP code, county, state) and various time ranges, facilitating nuanced analyses. Methods for data standardization from diverse sources were employed to enable seamless integration with other healthcare datasets. Comprehensive documentation was also developed to ensure usability and reproducibility. A systematic data update cycle was established to maintain relevance in the dynamic research landscape. These structured approaches collectively provide a reliable, scalable, and user-friendly tool for researchers to unlock critical health insights from our Exposome database.

## Development of the SDoH-EHR Linkage Tool

The integration of SDoH and environmental datasets with EHR has emerged as a crucial frontier in healthcare research. While EHR systems capture patient-level SDoH elements, community-level SDoH and environmental datasets provide essential contextual information that enriches research capabilities. This multi-level data integration enables sophisticated analyses in public health research, clinical pattern discovery, and temporal studies. To facilitate the seamless integration of community-level SDoH data with EHR systems, we developed a comprehensive data linkage framework. This framework consists of two primary components: a privacy-preserving local geocoding tool and a secure web-based linkage interface.

### Privacy-Preserving Geocoding Implementation

Patient privacy protection and HIPAA compliance are paramount in our implementation strategy. After extensive evaluation of available geocoding solutions, we selected

DeGauss as our core geocoding engine based on its robust privacy features and reliability. The geocoding process occurs entirely within the local institutional environment, ensuring that protected health information (PHI) never leaves secure institutional boundaries. The system outputs only de-identified coordinates and FIPS-11 codes, maintaining strict compliance with privacy regulations. For institutions utilizing the OMOP Common Data Model, we developed specialized integration scripts that streamline the geocoding process while maintaining compatibility with existing data structures. These scripts facilitate automated processing of address data while preserving the standardized OMOP format.

### Secure Web-Based Data Linkage

The second component of our framework comprises a secure web application that manages the linkage between de-identified location data and community-level SDoH information. As shown in Figure 2, this application is hosted within a protected infrastructure environment, implementing multiple security layers to prevent unauthorized access and potential system vulnerabilities. Access to the linkage system is managed through a registration-based portal (https://exposome.rc.ufl.edu/), ensuring appropriate authentication and authorization controls. We also developed a task tracking system for each user which will indicate the status of submitted linkage jobs and send notifications through emails. Comprehensive performance testing has validated the framework's capability to handle diverse data volumes efficiently. Our testing protocol included scenarios ranging from small-scale research projects to large institutional implementations, demonstrating consistent reliability and processing efficiency across different use cases.

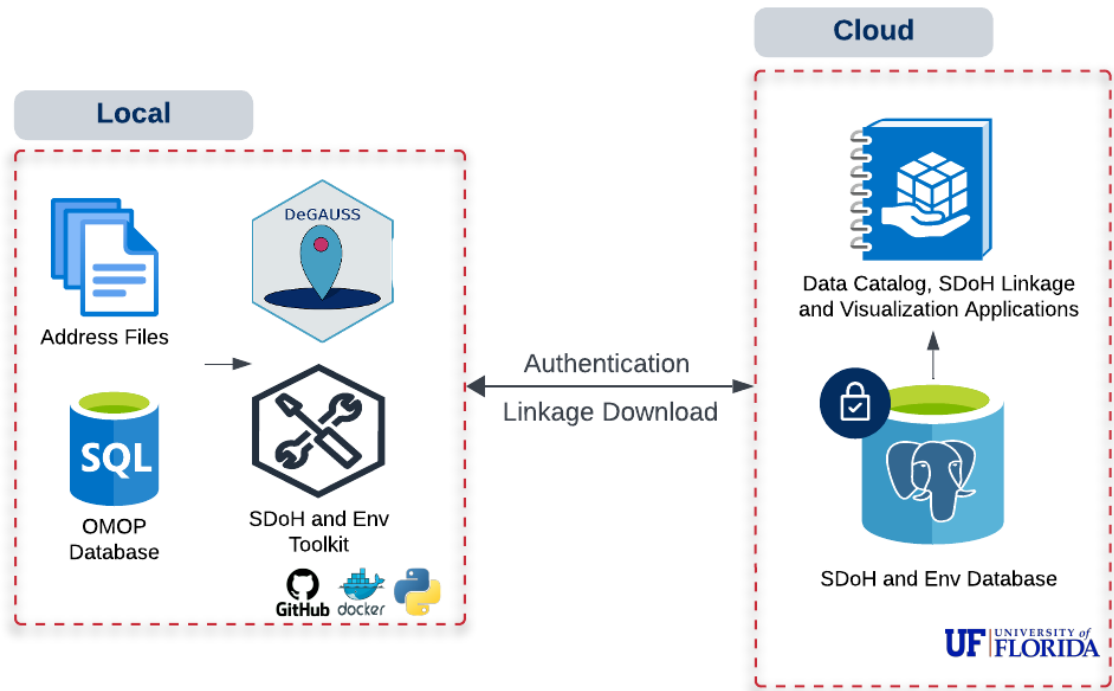

**Figure 2. A comprehensive framework for SDoH-EHR linkage.** This framework comprises two key components: a privacy-preserving local geocoding tool and a secure web-based linkage interface.

**Figure 3. A secure and user-friendly web-based SDoH<->EHR linkage interface.**

# Web Application for Visualization and Analysis

## Building the Web Application for the Exposome database

To enhance the usability and accessibility of the Exposome database, we have developed a comprehensive web application (Figure 3). This tool is designed to serve a wide user base, including researchers, clinicians, care managers, operations staff, and community partners, by centralizing SDoH data and making it more accessible through advanced tools and visualization techniques. The application addresses several key challenges by offering two primary features: 1) a database display that allows users to easily browse and review available datasets, ensuring transparency and understanding of the data, and 2) an automated linking tool that streamlines the process of connecting address-converted Federal Information Processing Standards (FIPS) codes to the Exposome database, enabling efficient retrieval of relevant information with minimal user effort.

## Web-Based Interface

Adopting a human-centered design approach, our Exposome Dashboard prioritizes usability, accessibility, and the needs of end-users. The application features a comprehensive data catalog that allows users to browse and explore SDoH datasets with ease, ensuring that users can validate and understand the datasets before utilization. The catalog provides clear classifications for each dataset, search capabilities, and descriptive metadata, allowing for easy navigation. Users can apply intuitive demographic and geographic filters, simplifying data usage by filtering by boundary type and year. Clicking on a database name redirects users to the source page, offering direct access to its origin and further details. Each database is listed with the available years and clicking on a specific year reveals a data dictionary that provides detailed descriptions for the original source. Helpful prompts explain the types of boundary types and other key information in the filter section, ensuring users have the necessary context for informed selections. These database and boundary type filters empower users to explore data effectively, understand factors impacting healthcare and research, and appreciate the geographic implications of these factors.

A notable frontend feature is the SDoH linkage functionality in the "Tool" section, which allows users to upload either a zipped CSV file or a standalone CSV file. Users can then select the desired SDoH type for linkage, such as "EJI," "FOODACCESS," "NEIGHBORHOODATLAS," "SDoH" (AHRQ database), or "SVI." The backend processes the uploaded data and selected SDoH type, providing formatted results conducive to subsequent research. This functionality significantly enhances user experience by facilitating targeted data exploration and providing clarity and guidance during data exploration and filtering.

## Frontend & Backend Infrastructure

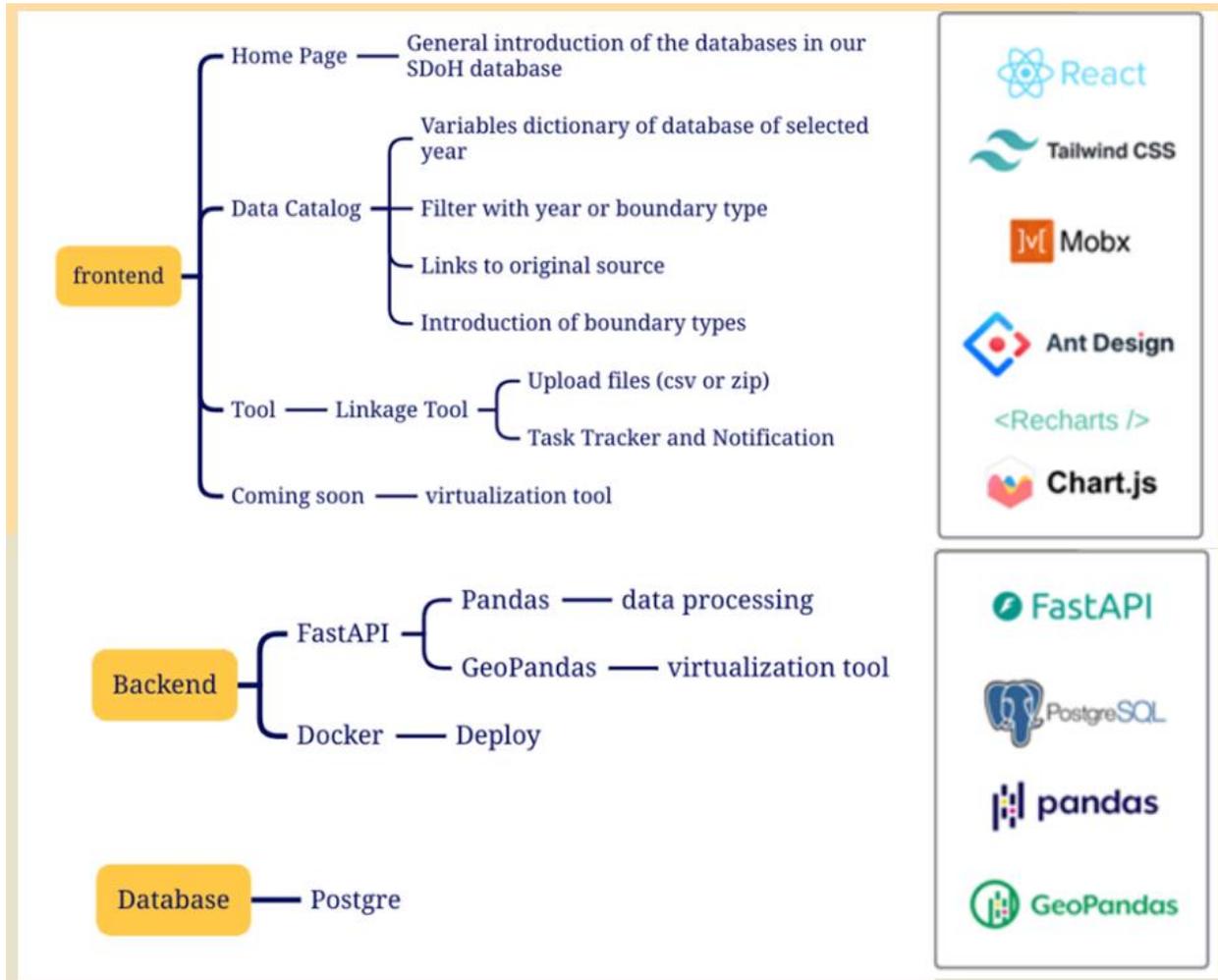

**Figure 4. Architecture of the SDoH web application, detailing frontend, backend, database components, and associated technologies**

As shown in Figure 4, the frontend development employs modern frameworks and libraries including React (https://react.dev/) for responsive user interfaces, MobX (https://mobx.js.org/README.html) for state management, Tailwind CSS (https://tailwindcss.com/) for styling, and Ant Design (https://ant.design/) for component design. This combination ensures a visually appealing and user-friendly interface.

The robust backend infrastructure, powered by FastAPI (https://fastapi.tiangolo.com/), efficiently handles complex data analyses, allowing for real-time insights and results. FastAPI's compatibility with libraries such as Pandas and GeoPandas enables seamless data manipulation and geospatial processing. Pandas is instrumental in processing substantial data necessary for linking user-uploaded data with addresses to SDoH datasets, while GeoPandas enhances the application's capability to manage spatial operations, streamlining workflows and improving the ability to process and analyze complex SDoH and environmental data effectively.

### Database Management

PostgreSQL has been chosen as the database management system for its superior performance in handling detailed SDoH and environmental data. It supports complex queries, indexing, and large datasets with ease. The PostGIS extension further enhances PostgreSQL's functionality by providing robust geospatial data storage, analysis, and querying capabilities, which are essential for managing geographical boundaries and spatial data. The reliability, scalability, and extensibility of PostgreSQL make it an ideal choice for managing the extensive and diverse Exposome database, supporting both real-time and historical analyses.

### Task Monitoring and Privacy

To ensure efficient task management and data security, the application includes a real-time dashboard for tracking the progress of tasks. Users receive updates on the status of the linkage process and notifications upon task completion. Additionally, the linked data is available for download for seven days, after which it is permanently deleted from the server, ensuring the security and privacy of user data. This modern, user-friendly architecture provides a streamlined and efficient experience for users, enhancing the overall utility of the Exposome database.

## Discussions

Previously, the World Health Organization (WHO) recognized disparate SDoH as factors driving health inequities, and worsened well-being and health outcomes. In a more recent 2024 report from the WHO further underscored the importance of accessing cross-sector, multi-level geographic data to effectively monitor SDoH in advancing health equity.[1, 22] Area-level SDoH have been investigated in wide-ranging population health fields[2] and their influence on health outcomes have been evidenced with studies done in various domains, such as aging[23] and cancer[7] research. SDoH indicating disadvantaged socioeconomic status were found to be associated with mortality in critical care patients.[24, 25] Postoperative complications increased with greater social vulnerability in pediatric and adult cohorts.[26, 27] Similarly, SDoH measures indicating adverse socioeconomic and neighborhood conditions were found to be influencing the health outcomes related to chronic[28] and acute illnesses[29]. Given the significant impact of SDoH on healthcare outcomes and research, we assembled Exposome, a large database coalescing and cataloging several datasets from distinct providers. Additionally, we shared an offline open-source geocoding toolkit with the community.

### Limitations

While the Exposome database and SDoH-EHR linkage tool presented in this paper represent significant advancements, several limitations should be acknowledged. First, the current database primarily includes area-level and neighborhood-level SDOH indicators, with limited individual-level data. Expanding the database to incorporate granular, individual-level measures will enhance the ability to capture nuanced relationships in SDoH outcomes. Additionally, the Exposome database is currently limited geographically. Scaling the database geographically will increase its generalizability and utility for researchers and healthcare providers nationally.

Furthermore, the SDoH-EHR linkage tool, while automated, still requires manual upload of the required address information. Developing more advanced, self-updating mechanisms for the linkage tool would help streamline the integration process and reduce the burden on users. Finally, future iterations of this project should explore ways to incorporate feedback from end-users, such as researchers and clinicians, to continuously improve the database and tool's functionality and user experience.

Despite these limitations, the Exposome database and SDoH-EHR linkage tool represent a valuable resource that addresses critical gaps in existing datasets and tools. The authors are committed to continuously enhancing this resource and welcome collaboration and feedback from the broader research community to unlock new health insights and address health disparities.

## Future Directions and Expansion

### Enhancements to the Exposome database

Future iterations of the Exposome database will incorporate additional indicators such as transportation access, food insecurity, and environmental hazards, providing a more comprehensive view of social determinants affecting health outcomes. The inclusion of longitudinal data will allow researchers to track trends and changes over time, offering insights into the long-term effects of social and environmental factors. Efforts to improve data quality and standardization will be continuous, focusing on refining data collection methods, enhancing validation processes, and ensuring compatibility with other health data standards. These improvements will facilitate seamless integration with additional datasets and enhance the reliability of research findings.

### Geospatial Visualization Tools

To further enhance the analytical capabilities of the system, we plan to develop geospatial visualization tools. These tools will provide users with intuitive means of analyzing the effects of different social determinants across various geographic regions. Features will include multi-factor mapping for comparison across different determinants and temporal analysis to track changes over time. By systematically integrating and visualizing geospatial data, users can gain deeper insights into spatial patterns, trends, and disparities, offering a comprehensive understanding of the impact of social determinants on specific populations. This will aid in the implementation of targeted public health interventions.

### Advanced Analytical Toolsets and AI Integration

We aim to develop advanced analytical tools, including machine learning models and predictive analytics, to leverage the rich SDoH and EHR data. These tools will predict health outcomes, identify at-risk populations, and model potential interventions. By incorporating machine learning, we can uncover complex patterns and relationships that traditional statistical methods might miss, enhancing our ability to forecast health trends and tailor interventions more precisely. AI integration will also enhance the interpretation of complex datasets, providing more accurate and actionable insights, thereby supporting researchers in identifying critical factors influencing health outcomes and addressing health equity. This approach will be validated through collaborations, such as those with NorthShore University Health System, which have demonstrated the effectiveness of AI, particularly Natural Language Processing (NLP), in extracting actionable insights from extensive datasets.

### Long-Term Vision

Our long-term vision includes fostering continued collaboration with the research community to maintain the relevance and utility of the database and tools. We plan to establish partnerships with public health organizations, healthcare providers, and academic institutions to broaden the impact of our toolset. Ensuring the sustainability and scalability of the database and tools is a key priority, which includes securing funding for ongoing development, expanding the user base, and continuously updating the database to reflect the latest research and data trends. Additionally, we envision expanding the database to include global health data, enabling comparative studies across different countries and regions and providing a broader perspective on the impact of social determinants on health. This global expansion aims to foster innovative approaches to health research and policy development, contributing to a more comprehensive understanding of health determinants worldwide.

## Conclusion

The development of the Exposome database and the SDoH-EHR linkage tool represents a significant advancement in the field of health disparities research. By providing a comprehensive, open-access resource that integrates multidimensional social and environmental data with EHR, we have created a powerful tool for researchers and healthcare providers. This resource addresses critical gaps in existing databases and tools, offering a more granular and standardized approach to analyzing the impact of social determinants on health outcomes. The freely accessible nature of these resources ensures that a wide range of stakeholders, including researchers, clinicians, and public health professionals, can leverage the data to drive innovative research and inform evidence-based interventions. The integration of SDOH data into healthcare delivery systems has the potential to transform patient care by enabling more personalized and contextually informed treatment plans. Looking ahead, the continuous expansion and enhancement of the database, coupled with the development of advanced analytical tools and AI integration, will further empower the research community. These efforts will facilitate deeper insights into the complex interplay between social determinants and health, ultimately contributing to improved health equity and outcomes. In conclusion, the Exposome database and linkage tool are poised to become indispensable assets in the quest to understand and address health disparities. By fostering collaboration and providing robust, scalable resources, we aim to support the ongoing efforts to achieve a more equitable healthcare system.

## Acknowledgments

A.B. was supported by R01 GM110240 from the National Institute of General Medical Sciences (NIH/NIGMS), R01 DK121730 from the National Institute of Diabetes and Digestive and Kidney Diseases (NIH/NIDDK), OT2 OD032701 from the National Institutes of Health Bridge2AI Program, and the University of Florida President's

Strategic Initiatives, a program sponsored by the State of Florida.  T.O.B. was supported by R01 GM110240 from the National Institute of General Medical Sciences (NIH/NIGMS), R01 DK121730 from the National Institute of Diabetes and Digestive and Kidney Diseases (NIH/NIDDK), by OT2 OD032701 from the National Institutes of Health Bridge2AI Program. This research was funded by R01 GM110240 from the National Institute of General Medical Sciences (NIH/NIGMS). The funding sources had no role in the design and conduct of the study; collection, management, analysis, and interpretation of the data; preparation, review, or approval of the manuscript; and decision to submit the manuscript for publication. The content is solely the responsibility of the authors and does not necessarily represent the official views of the National Institutes of Health and other funding sources.